\documentclass[10pt,prd,twocolumn,eqsecnum,amssymb,amsmath,showpacs,a4paper,superscriptaddress,nofootinbib,floatfix]{revtex4-2}

\usepackage{units,tabularx,graphicx,dcolumn,bm,tikz,soul,geometry,hyperref,braket}

\usepackage{xcolor}
\usepackage{float}

\newcommand{\sx}{\mathsf{x}}

\newcommand{\ts}[1]{ _{\text{#1}} }
\newcommand{\enclose}[1]{\left(#1\right)}
\newcommand{\benclose}[1]{\left[#1\right]}

\makeatletter
\let\cat@comma@active\@empty
\makeatother

\usepackage{xcolor}
\hypersetup{  colorlinks,
              linktoc         = page, 
              urlcolor        = {green!80!blue},
              linkcolor       = {orange!60!black}, 
              urlcolor        = {cyan!80!black}, 
              citecolor       = {cyan!80!black}, 
              anchorcolor     = {yellow}
}

\begin{document}

\title{Superposed circular motion Unruh effect in (3+1) dimensions}






\author{Taylor Cey}
\affiliation{Department of Physics and Astronomy, University of Waterloo,
200 University Ave W, Waterloo, Ontario N2L 3G1, Canada}
\author{Cisco Gooding}
\affiliation{Laboratoire Kastler Brossel, Sorbonne Universit\'{e}, CNRS, ENS-Universit\'{e} PSL,  Coll\`{e}ge de France, 4 place Jussieu, F-75252 Paris, France}
\author{Robert Mann}
\affiliation{Department of Physics and Astronomy, University of Waterloo,
200 University Ave W, Waterloo, Ontario N2L 3G1, Canada}
\affiliation{Perimeter Institute for Theoretical Physics, 31 Caroline St. N., Waterloo, Ontario N2L 2Y5, Canada}
\affiliation{Department of Applied Mathematics, University of Waterloo,
200 University Ave W, Waterloo, Ontario N2L 3G1, Canada}
\affiliation{
Institute for Quantum Computing, University of Waterloo,
200 University Ave W, Waterloo, Ontario N2L 3G1, Canada}
\date{\today}

\begin{abstract}
Using a recently-introduced quantum control model for Unruh-DeWitt detectors in superpositions of classical trajectories, we investigate the response of a detector interacting with a massless scalar quantum field in (3+1) dimensions along a superposition of circular trajectories. We present numerical results for the transition probability and effective temperature of such a detector in four distinct geometric scenarios: (a) concentric, vertically-stacked trajectories, (b) planar, horizontally-displaced trajectories, (c) static central point and surrounding circular trajectory, and (d) concentric, planar circular trajectories. For Gaussian switching functions that are much broader than the acceleration timescale, in case (a) we find only minor deviations from the well-known, effectively thermal response of a single circular trajectory, whereas in case (c) we find a significant reduction in the effective temperature and greater variation with energy gap. We conclude with a discussion of a potential analogue implementation in ultracold atom systems.
\end{abstract}

\maketitle

\section{Introduction}

One of the central tools to probe the structure of relativistic quantum field theories is the Unruh-DeWitt (UDW) model \cite{Unruh1979evaporation,DeWitt1979}, which describes a particle detector interacting locally with a quantum field. In its original form, the UDW model was used to demonstrate the Unruh effect: a uniformly accelerated detector coupled to a scalar vacuum field responds thermally, with the associated temperature varying linearly with the proper acceleration of the detector \cite{Unruh1979evaporation}. 

Recently there has been renewed interest in UDW detectors undergoing uniform circular motion, due to a proposed analogue realization in laser-coupled ultracold atom systems \cite{Analog2,Analog1,Analog4}. Similar to the case of linear acceleration, a  UDW detector traveling along a circular worldline exhibits an approximately thermal response, which also varies linearly with the proper acceleration \cite{Unruh1998}. This proposed analogue realization opened the door for the experimental exploration of other relativistic quantum information phenomena, such as entanglement harvesting \cite{Analog3}.

There is also an emerging interest in extending the study of UDW detector response along an individual classical trajectory to allow superpositions of trajectories \cite{Foo:2020xqn,Foo:2020dzt}. Accelerated trajectories and questions of effective thermality are of particular interest \cite{PhysRevD.102.045002,Wood:2021xmw,Foo:2023yxj}, as are the applications for related operational inquiries in quantum gravity
\cite{Foo:2020jmi,Foo:2021gkl,Foo:2021exb,Foo:2022dnz,Suryaatmadja:2023onb}. A preliminary proposal has been suggested by the authors to test the trajectory superposition formalism \cite{twoplusone_preprint}, as a first step in providing experimental guidance for such endeavours.


Our paper is organized as follows. Section \ref{sec:UDW} introduces the model of a quantum-controlled UDW detector interacting with a scalar (3+1)-dimensional quantum field. In Section \ref{sec:circ}, uniform circular trajectories are introduced, and four distinct scenarios for superpositions of circular trajectories are analyzed. Section \ref{sec:Teff} then presents results on the effective temperature associated with these four scenarios, with comparisons made to the case of incoherent (classical) mixtures of trajectories. Finally, we conclude with a discussion of the results and implications for experimental implementation in Section \ref{sec:Disc}.

\section{Unruh-DeWitt Detectors}
\label{sec:UDW}

We construct a system involving a massless scalar quantum field $\hat{\phi}$ in a (3+1) Minkowski spacetime and a single pointlike Unruh-DeWitt (UDW) detector with a ground $\ket{g}$ and excited state $\ket{e}$. Additionally, we introduce a quantum control degree of freedom that dictates the trajectory of our detector $\ket{\psi}$. 

We initialize our field and detector system in their respective ground states. Our total system is then given by
\begin{equation}
    \ket{\Psi} = \ket{g} \otimes \ket{\psi} \otimes \ket{0}
\end{equation}
where $\ket{0}$ denotes the vacuum state of the $\hat{\phi}$ field and the specific trajectory taken by the detector through spacetime is determined by a control state in an equal, coherent superposition of $N$ trajectories:
\begin{equation}
    \ket{\psi} = \frac{1}{\sqrt{N}}\sum_{j=1}^N \ket{j}.
\end{equation}
We will restrict attention in this work to $N=2$. The corresponding density operator is then
\begin{equation}
    \rho\ts{control} = \ket{\psi}\bra{\psi}=\frac{1}{2}\sum_{i=1}^2\sum_{j=1}^2 \ket{i}\bra{j}
\end{equation}
In contrast, an incoherent (classical) mixture of trajectories is described by the density operator
\begin{equation}
    \rho\ts{control,mixed} = \frac{1}{2}\sum_{j=1}^2 \ket{j}\bra{j}\, .
\end{equation}

In the interaction picture, the Hamiltonian describing our interaction is given by
\begin{align}
    \hat H ( \tau)
    &=
        \lambda\hat \mu(\tau) \sum_{j=1}^2 \eta_j(\tau) 
        \hat \phi(\mathsf{x}_j(\tau))\otimes \ket{j}\bra{j},
\end{align}
where $\tau$ is the proper time along a trajectory $x_j(\tau)$, and $\eta_j(\tau)=\exp(-\tau^2/2\sigma^2)$ is a Gaussian switching function. Here, $\hat \mu_j(\tau) $ is the monopole moment given by
\begin{align}
    \hat \mu_j(\tau) 
    &=
        \ket{e} \bra{g} e^{ i \Omega \tau }
        +
        \ket{g} \bra{e} e^{ -i \Omega \tau },
\end{align}
which describes the dynamics of the UDW detector with the field along it's trajectory, where $\Omega$ is the energy gap of the detector. When $\Omega<0$, the initial state $\ket{g}$ is of a higher energy than the secondary state $\ket{e}$; hence, in this case the roles of these two states are reversed compared the $\Omega>0$ case. After tracing out the field, the final state of the detector (to leading order in $\lambda$) is given by
\begin{equation}
    \hat\rho_D = 
    \begin{bmatrix}
        1-P_E & 0\\
        0 & P_E
    \end{bmatrix}
\end{equation}
where the excitation probability is given by
\begin{equation}
    P_E = \frac{\lambda^2}{4}\sum_{i=1}^2 \sum_{j=1}^2 P_{ij},
    \label{eq:Psum}
\end{equation}
and
\begin{align}\label{eq:TransitionP_ij}
    &P_{ij} = \\
    &\int d\tau_i\int d\tau_j'\: \eta(\tau_i)\eta(\tau_j') e^{-i\Omega(\tau_i-\tau_j')} W_{ji}(\tau_i,\tau_j').\nonumber
\end{align}
This gives rise to non-local contributions of the Wightman function when $i\neq j$, and therefore possible self-interference effects of the UDW detector along its different trajectories. The Wightman function here is pulled back along the trajectories of the detector,
\begin{equation}
    W(\mathsf{x}, \mathsf{x}') = \bra{0} \hat\phi(\mathsf{x}(\tau)) \hat \phi(\mathsf{x}'(\tau')) \ket{0}\,
\end{equation}
and is given by \cite{birrell1984quantum}
\begin{align}
    W_{ji}(\tau_i, \tau_j')
    &=
        \lim_{\epsilon\rightarrow0}\dfrac{-1/4\pi^2}{ (t_i-t_j'-i \epsilon)^2 - (x_i - x_j')^2 }\,,
    \label{eq:wightman_def}
\end{align}
where the Wightman function is understood to be a function of the trajectories, and therefore of the proper times. The coordinates given above are given as functions of the proper times such that $x_i=x_i(\tau_i)$ and $t_i = t_i(\tau_i)$. The $\epsilon$ given in equation \eqref{eq:wightman_def} is to be interpreted in the distributional sense, where $\epsilon\rightarrow 0^+$; we will omit this quantity in some subsequent expressions for brevity. When the Wightman function depends only on $s=\tau-\tau'$, equation \eqref{eq:TransitionP_ij} takes the form
\begin{equation}
    P_{ij} = \sigma\sqrt{\pi}\int ds\: e^{-\frac{s^2}{4\sigma^2}}e^{-i\Omega s} W_{ji}(s)\,.
    \label{eq:StationaryP_ij}
\end{equation}

\begin{figure*}[t]
\centering
\includegraphics[width=\textwidth]{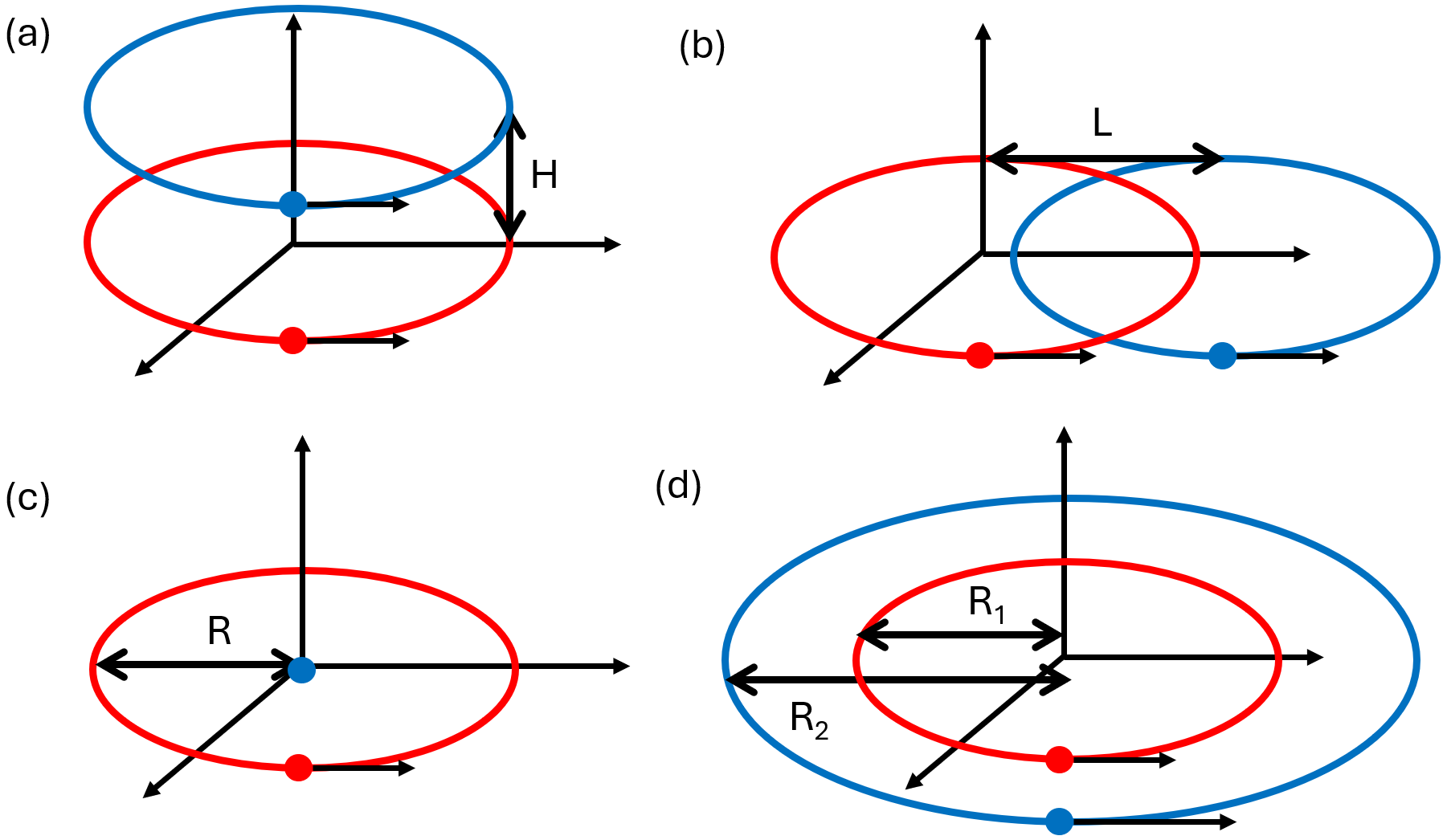}
\caption{
    The superposed trajectories of the detector for a variety of configurations. (a) Vertically displaced, (b) horizontally displaced, (c) static and (d) concentric.
}
\label{fig:trajectories}
\end{figure*}

\section{Circular Trajectories}
\label{sec:circ}

When defining our system we must first select trajectories for which the detector will travel in superposition. In principle, we can select any number of trajectories for the UDW detector to be in superposition, but here we select two trajectories for simplicity. For our purposes, we look at two distinct circular trajectories in a variety of configurations. Unless otherwise stated, the radii and angular velocities of the two circular trajectories are assumed to be the same.

We begin by reviewing results for a single circular trajectory. A standard form for a circular trajectory is given by
\begin{align}
    \sx(\tau)
    &=
        \enclose{
            \gamma \tau, 
            R\cos (\omega \gamma \tau),
            R\sin (\omega \gamma \tau),
            0
        }\,.
\end{align}
This gives a Wightman function of the form
\begin{align}
    W_{jj}(s) 
        = 
            -\dfrac{1}{4\pi^2}
            \dfrac{1}{\gamma^2 s^2 - 4 R^2\sin^2(\omega\gamma s/2)}\, ,
    \label{eq:diagonal_wightman}
\end{align}
where $\omega$ is the angular frequency of the detector along the circle, $R$ is the radius of the circle, $\gamma = 1/\sqrt{1 - \omega^2R^2}$ is the usual Lorentz factor, and $s := \tau - \tau'$. Note also that the tangential speed is $v=R\omega$ and the proper acceleration has magnitude $a=\gamma^2\, \omega^2 R$. Alternatively, these expressions can be inverted, such that $R=v^2/a(1-v^2)$ and $\omega=a(1-v^2)/v$.

A useful reference to keep in mind is the response of a static detector with an infinite-duration interaction with the field. In this simple case, the Wightman function is given by
\begin{align}\label{staticWightman}
    W(s)=-\frac{1}{4\pi^2}\frac{1}{(s-i\epsilon)^2}\, ,
\end{align}
and the transition rate is proportional to the response function
\begin{align}
    \mathcal{F}(\Omega)=\int_{-\infty}^\infty ds\, e^{-i\Omega s}\,W(s)\, .
\end{align}
Straightforward contour integration yields 
\begin{align}\label{staticresponse}
    \mathcal{F}(\Omega)=-\frac{\Omega}{2\pi}\Theta(-\Omega)\, .
\end{align}
Hence, a completely static detector only transitions when initialized in the excited state, whereas we will see that an accelerated detector gains the ability to be spontaneously excited. This is an aspect of the well-known linear-motion Unruh effect, and occurs also for circular trajectories. The rest of  our investigation is dedicated to examining the effects of a superposition in contrast to the standard results one finds with single trajectories.

In this Section, we focus on the transition probability for detectors initialized in the ground state $\ket{g}$, with $\Omega>0$. The attention is therefore restricted to the excitation spectrum; in the following Section \ref{sec:Teff}, we will consider the effective temperature, which involves both the excitation ($\Omega>0$) and de-excitation ($\Omega<0$) spectra.

\subsection{Vertically displaced circles}
\label{sec:vert_circ}

The first orientation of superposed trajectories of interest is circles stacked vertically on top of one another, which corresponds to (a) in Figure \ref{fig:trajectories}. We assume that the two trajectories are aligned throughout the motion such that the coordinates are given by
\begin{align}
    \sx_1(\tau)
    &=
        \enclose{
            \gamma \tau, 
            R\cos (\omega \gamma \tau),
            R\sin (\omega \gamma \tau),
            0
        }\,\\
    \sx_2(\tau)
    &=
        \enclose{
            \gamma \tau, 
            R\cos (\omega \gamma \tau),
            R\sin (\omega \gamma \tau),
            H
        }\,.
\end{align}
The Wightman function between the two paths then becomes
\begin{equation}\label{eq:WightmanVertical}
    W\ts{v}(s) = -\frac{1}{4\pi^2}\frac{1}{\gamma^2s^2 - 4R^2\sin^2\enclose{\frac{1}{2}\omega\gamma s} - H^2}\,.
\end{equation}
The simple form of the Wightman is maintained thanks to the stationarity of the two trajectories, meaning that the angle of the velocity vector of the detector and the spatial displacement vector between the trajectories remains constant throughout the motion. 

\begin{figure}
\centering
\includegraphics[width=\columnwidth]{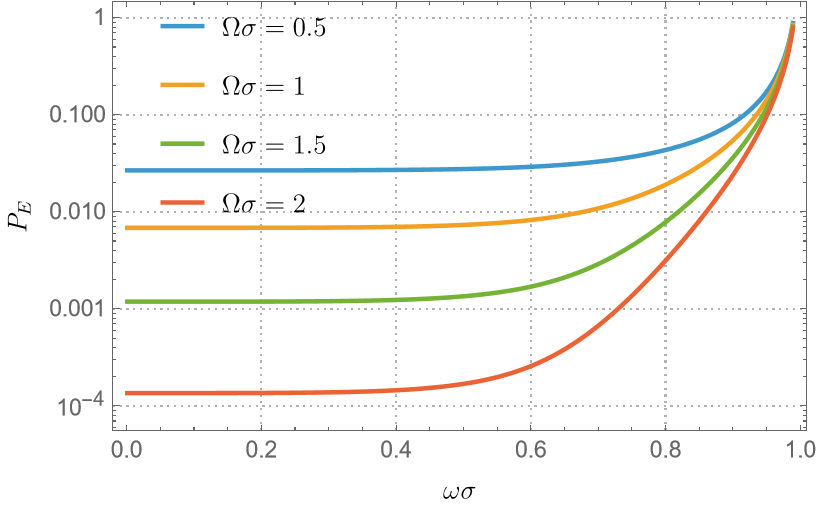}
\caption{
    The transition probability of a detector in a  superposition of vertically displaced circles versus rotational frequency, $\omega\sigma$. Here $H/\sigma=1$ and $R/\sigma=1$.
}
\label{fig:Zgap_P_vs_omega}
\end{figure}

\begin{figure}[t]
\centering
\includegraphics[width=\columnwidth]{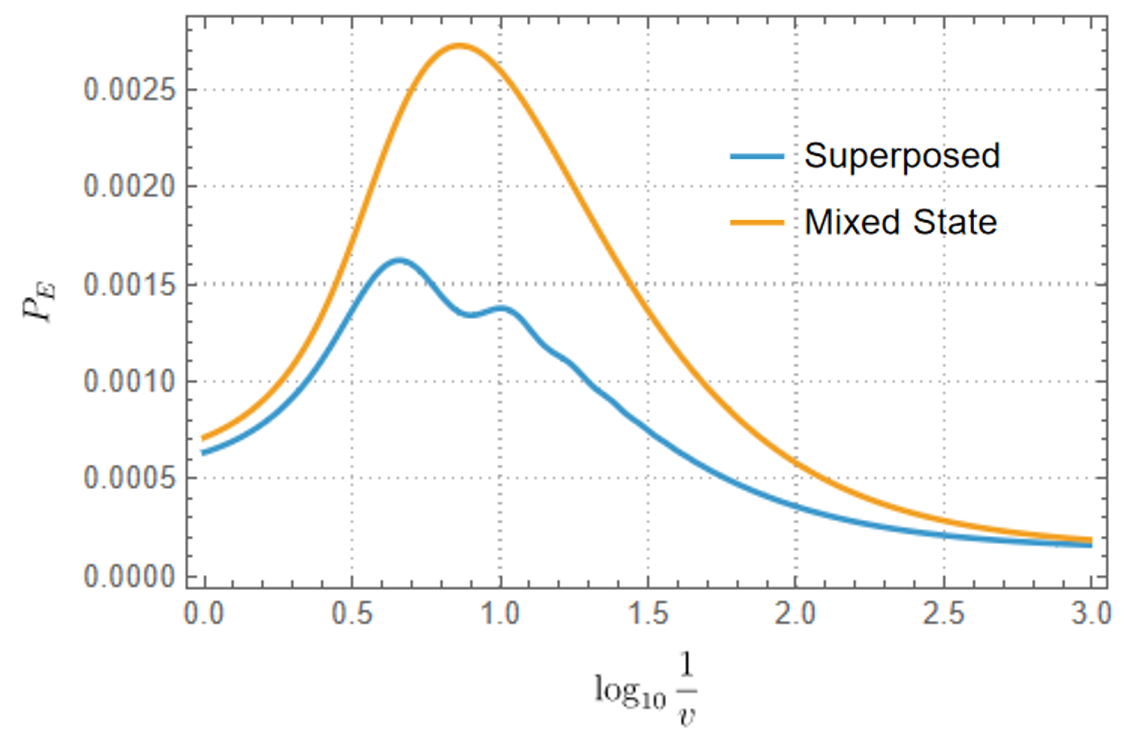}
\caption{
    The transition probability for an ordinary non-superposed detector traveling in a circular trajectory versus the transition probability of a detector in a superposition of two identical trajectories displaced vertically by a distance $H/\sigma=1$. In both cases $a\sigma = 1$ and $\Omega\sigma=2$.
}
\label{fig:Zgap_mixedcomparison}
\end{figure}

\begin{figure*}[t]
\centering
\includegraphics[width=\textwidth]{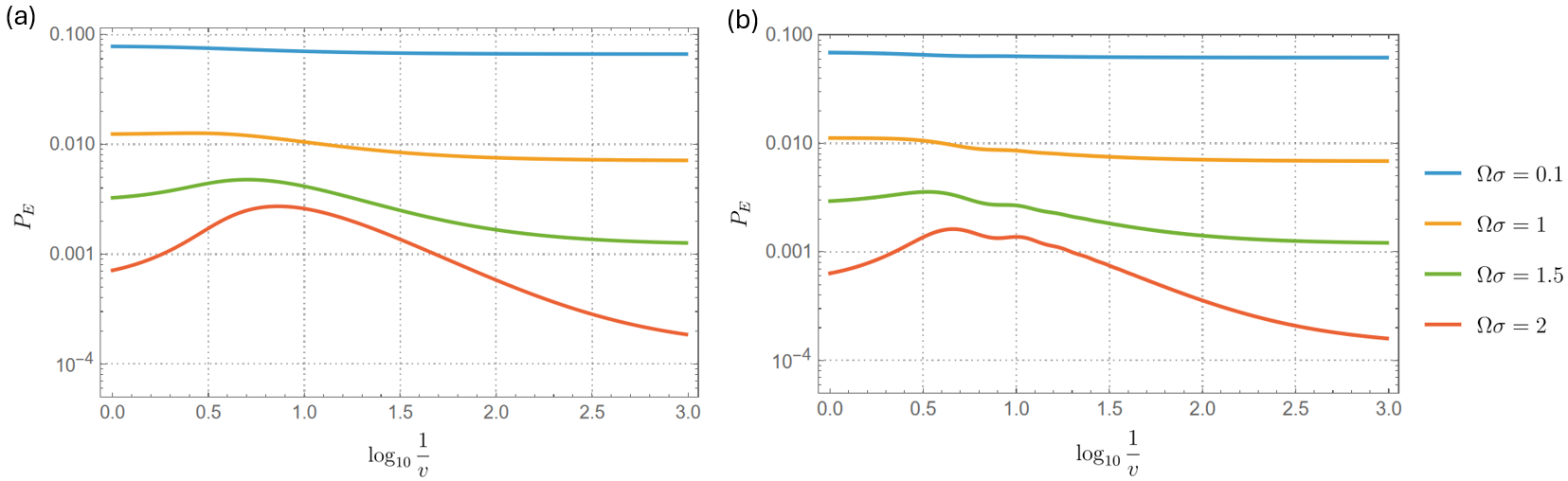}
\caption{
    (a) The transition probability of a single detector as a function of the tangential velocity of the detector. Here we hold the centripetal acceleration of the detector to be constant at $a\sigma=1$. (b) The transition probability of the superposed detector in a vertically stacked configuration for the same parameters, at a distance apart of $H/\sigma=1$.
}
\label{fig:Zgap_P_vs_logv}
\end{figure*}

The case of vertically displaced circles is of special interest for numerical computation due to the fact that the trajectories are stationary with respect to one another, because this allows the Wightman function \eqref{eq:WightmanVertical} to depend only on the difference in proper time between the trajectories. Not only does this make for easier computation, but we expect (based on previous work \cite{Bozanic:2023}) that  the behavior of the detector  will differ compared to nonstationary examples.

Figure \ref{fig:Zgap_P_vs_omega} shows the transition probability for a superposition of trajectories versus angular velocity. We see that as the velocity of the detector is increased in both trajectories, there is a corresponding increase in the response of the detector. This is an expected result, since the circular motion Unruh effect implies that the response of a detector should increase along with its centripetal acceleration. However, this is not unique to a superposition of trajectories and the contribution of the interference terms $i\neq j$ to the response are suppressed by the rapid increase in the ordinary excitation $i=j$ terms in the sum in equation \eqref{eq:Psum}. 

To better observe the effects of the superposition on the detector, we can instead vary the linear parameters of the circular trajectory and look at regimes where the centripetal acceleration $a$ is held constant. Here we expect the response of the detector to not be so heavily dominated by the ordinary excitation terms. Figure \ref{fig:Zgap_mixedcomparison} shows a comparison between the response function of the superposed detector to that of the mixed state case, which in this case is equivalent to a detector traveling along a single trajectory. We see that at constant acceleration, the 
transition probability in the 
superposed case has a significantly dampened peak. Interestingly we observe that the superposition has the effect of introducing an oscillation in the probability specifically over the range of velocities in the vicinity of the peak of the response in the mixed case.

Figure \ref{fig:Zgap_P_vs_logv} shows the same comparison for several different energy gaps. The effects of  superposition are consistently strongest at the peak of the response for a specific range of large velocities (roughly $0.06<v<0.3$ in natural units). The interference effects of   superposition on the probability does not increase proportional to the increase in the overall response of the detector for smaller energy gaps, much like the peak. In contrast, the detector response is nearly identical for both near relativistic and slow velocities.

\begin{figure}
\centering
\includegraphics[width=\columnwidth]{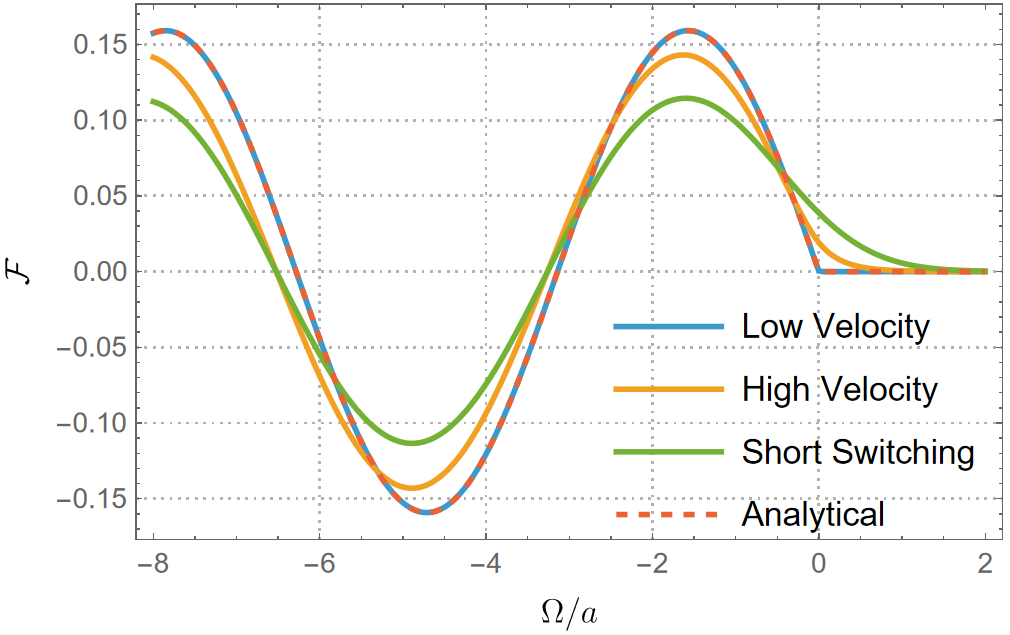}
\caption{
    The off-diagonal response function of a detector in a superposition of vertically displaced circular trajectories versus the energy gap of the detector, normalized by the centripetal acceleration. We set $\gamma=100$ to be in the ultrarelativistic regime, $Ha=1$, and $\sigma a=100$ is taken to be large to remove as much of the effects of the switching function as possible (orange curve). The low velocity curve (blue) is plotted with the same parameters except that $\gamma = 1.0001$ to fall well into the opposite regime. The short switching curve (green) is plotted for $\sigma=1$. We see the closest agreement between the low velocity numerical result and the analytical approximation \eqref{stackedultra0} (dashed red).
}
\label{fig:Zgap_UltraRel_analyticalcomparison}
\end{figure}

In the ultrarelativistic limit (i.e. $\gamma\rightarrow \infty$), the off-diagonal Wightman function behaves as
\begin{align}\label{stacked_asymp}
    W(s)\approx-\frac{1}{4\pi^2}\frac{1}{(s-i\epsilon)^2-H^2}\, .
\end{align}
The Wightman function \eqref{stacked_asymp} has poles at $s=i\epsilon\pm H$; for an interaction of infinite duration with constant switching function,  contour integration in this case leads to the response function
\begin{align}\label{stackedultra0}
    \mathcal{F}(\Omega)\approx-\frac{\Omega}{2\pi}\text{sinc}(\Omega H)\Theta(-\Omega)\, .
\end{align}
Note that the ultrarelativistic limit is taken such that $v\rightarrow 1$ at fixed $a$. Coincidentally, \eqref{stacked_asymp}, and therefore also \eqref{stackedultra0}, are the asymptotic expressions in the nonrelativistic limit ($\gamma\rightarrow 1$ at fixed $a$) as well. From \eqref{stackedultra0} we can deduce that the interference will become negligible compared to the inertial contribution to the diagonal response functions when $\Omega H \gtrsim \pi$ (equivalently, for vertical separations $H\gtrsim \pi/\Omega$); moreover, interference will also be suppressed for Gaussian switching when $\sigma$ is sufficiently less than $H$. Figure \ref{fig:Zgap_UltraRel_analyticalcomparison} shows a comparison of the asymptotic off-diagonal response function \eqref{stackedultra0} with numerical calculations of the response using the exact off-diagonal Wightman \eqref{eq:WightmanVertical} in both the ultrarelativistic and nonrelativistic limits. Further comparison is made with the response for a shorter switching function. 

\subsection{Horizontally displaced circles}

The horizontally displaced circles are defined by trajectories that are separated in the plane of rotation, shown in Figure~\ref{fig:trajectories} (b). An example of such a trajectory is given by
\begin{align}
    \sx_1(\tau)
    &=
        \enclose{
            \gamma \tau, 
            R\cos (\omega \gamma \tau),
            R\sin (\omega \gamma \tau),
            0
        }\,\\
    \sx_2(\tau)
    &=
        \enclose{
            \gamma \tau, 
            R\cos (\omega \gamma \tau) + L,
            R\sin (\omega \gamma \tau),
            0
        }\,.
\end{align}
Unlike the previous case, this set of trajectories is not stationary since   the displacement 
vector between the two superposed     detector positions  is in the plane  of   motion. The Wightman function between the trajectories then becomes
\begin{widetext}
\begin{equation}
    W\ts{h}(s,p) = -\frac{1}{4\pi^2}\frac{1}{\gamma^2s^2 - 4R^2\sin^2\enclose{\frac{1}{2}\omega\gamma s} - L^2 - 4LR\sin\enclose{\frac{1}{2}\omega\gamma s}\sin\enclose{\frac{1}{2}\omega\gamma p}} 
\end{equation}
\end{widetext}
where $p = \tau + \tau'$, signatory of the loss of   stationarity.

\begin{figure}
\centering
\includegraphics[width=\columnwidth]{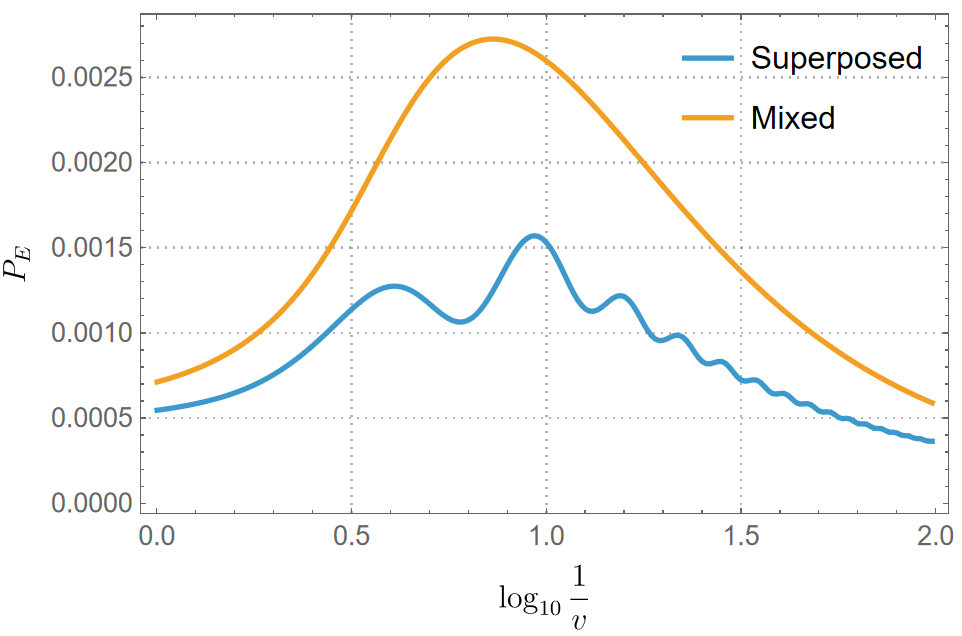}
\caption{
    (Blue) The transition probability of a detector in a superposition of horizontally displaced circular trajectories versus tangential velocity. (Orange) The transition probability for an equivalent mixed state system of identical trajectories. Here $\Omega\sigma=2$ and the centripetal acceleration of the circles is $a\sigma=1$.
}
\label{fig:Hgap_mixedcomparison}
\end{figure}

Figure \ref{fig:Hgap_mixedcomparison} shows a comparison of the transition probability for the detector in superposition versus the mixed state case or equivalently a single path response. Comparing this figure with figure \ref{fig:Zgap_mixedcomparison} shows that much of the effect of the superposition remains the same, such as the overall reduced probability and large oscillations in probability near the peak. The interference effects of the superposition in this configuration die off far more slowly for low velocities.

\begin{figure}
\centering
\includegraphics[width=\columnwidth]{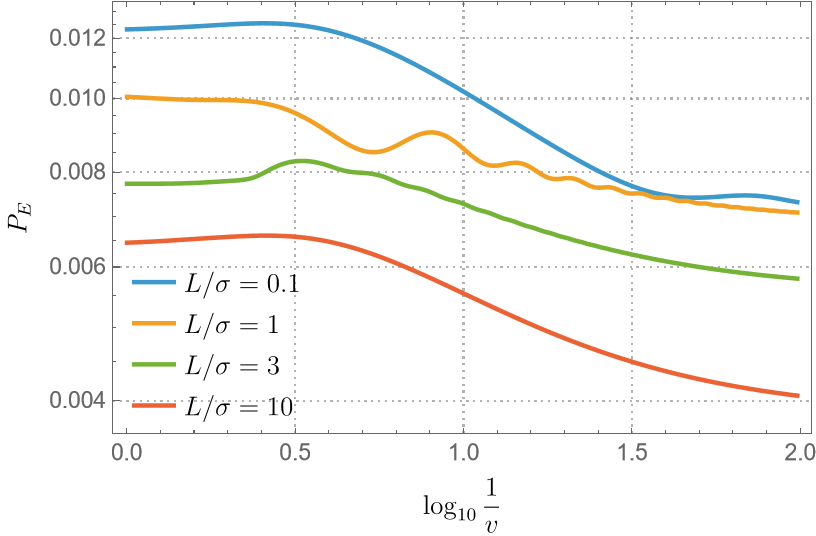}
\caption{
    The transition probability of a detector in a  superposition of horizontally displaced circles versus tangential velocity, $v$ for various displacements $L$ of their centres. Here $a=1$ and $\Omega=1$.
}
\label{fig:Hgap_P_Lcomparison}
\end{figure}

The nature of the oscillatory effect of the interference is strongly dependent on the displacement between the trajectories, as shown in figure \ref{fig:Hgap_P_Lcomparison}. For sufficiently separated trajectories, such as $L/\sigma=10$, the excitation probability takes the same form as a single detector and displays no oscillation whatsoever. As the displacement between the circles is reduced, the interference effect becomes much stronger along with the overall amplitude of the response.


\subsection{Static point}

One choice of trajectory of particular interest is that of a detector in a superposition of a static point in space and a circular trajectory orbiting that point, which corresponds to (c) in Figure \ref{fig:trajectories}. Such a trajectory can be described simply by
\begin{align}
    \sx_1(\tau)
    &=
        \enclose{
            \gamma \tau, 
            R\cos (\omega \gamma \tau),
            R\sin (\omega \gamma \tau),
            0
        }\,\\
    \sx_2(\tau)
    &=
        \enclose{
            \tau, 
            0,
            0,
            0
        }\,,
\end{align}

yielding the associated Wightman function

\begin{equation}
    W\ts{s}(\tau,\tau') = -\frac{1}{4\pi^2}\frac{1}{(\tau - \gamma \tau')^2 - R^2}.
    \label{eq:static_wightman}
\end{equation}

The Wightman function pulled back along only the world line of the stationary trajectory also differs such that

\begin{equation}
    W_{22}(s) = -\frac{1}{4\pi^2}\frac{1}{s^2}.
\end{equation}

\begin{figure}
\centering
\includegraphics[width=\columnwidth]{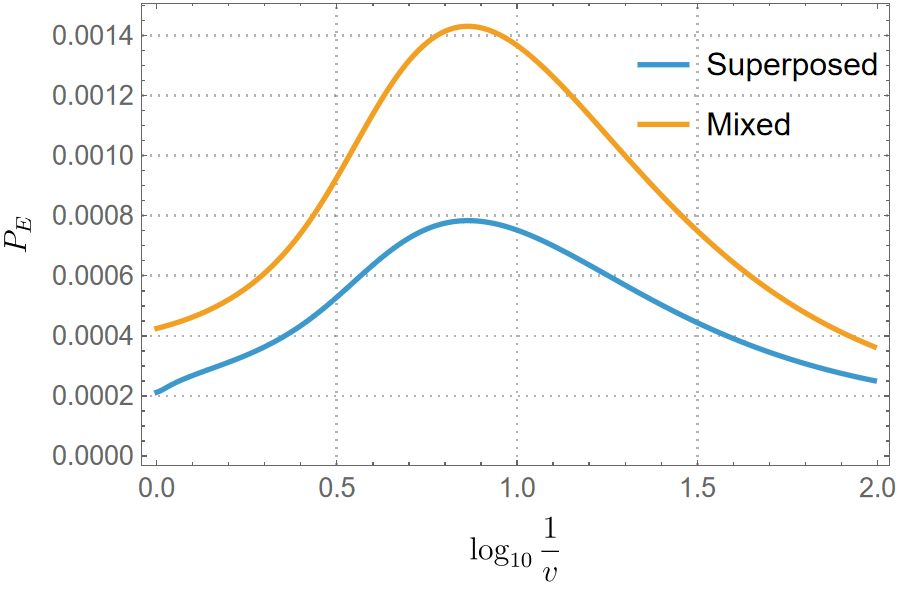}
\caption{
    (Blue) The transition probability of a detector in a superposition of a static point and an orbiting circular trajectory versus the tangential velocity of the orbit. (Orange) The transition probability for an equivalent mixed state system of identical trajectories. Here $\Omega\sigma=2$, and the centripetal acceleration of the orbit is $a\sigma=1$.
}
\label{fig:Static_mixedcomparison}
\end{figure}

This case presents a superposition of an accelerating trajectory and a non-accelerating trajectory. Figure \ref{fig:Static_mixedcomparison} shows how such a superposition compares to the classical mixed state for an equivalent system. Unlike the previous two configurations, this incoherent mixture does not correspond to a detector following a single trajectory because the two trajectories in superposition are no longer physically identical except for the displacement and instead are completely different. Unlike the previous two cases, we observe no oscillation in the response over a range of velocities. A superposition between an accelerating and non-accelerating trajectory results only in a damping effect. This is unsurprising given that a static detector has no probability for excitation under normal circumstances.

\begin{figure*}[ht]
\centering
\includegraphics[width=\textwidth]{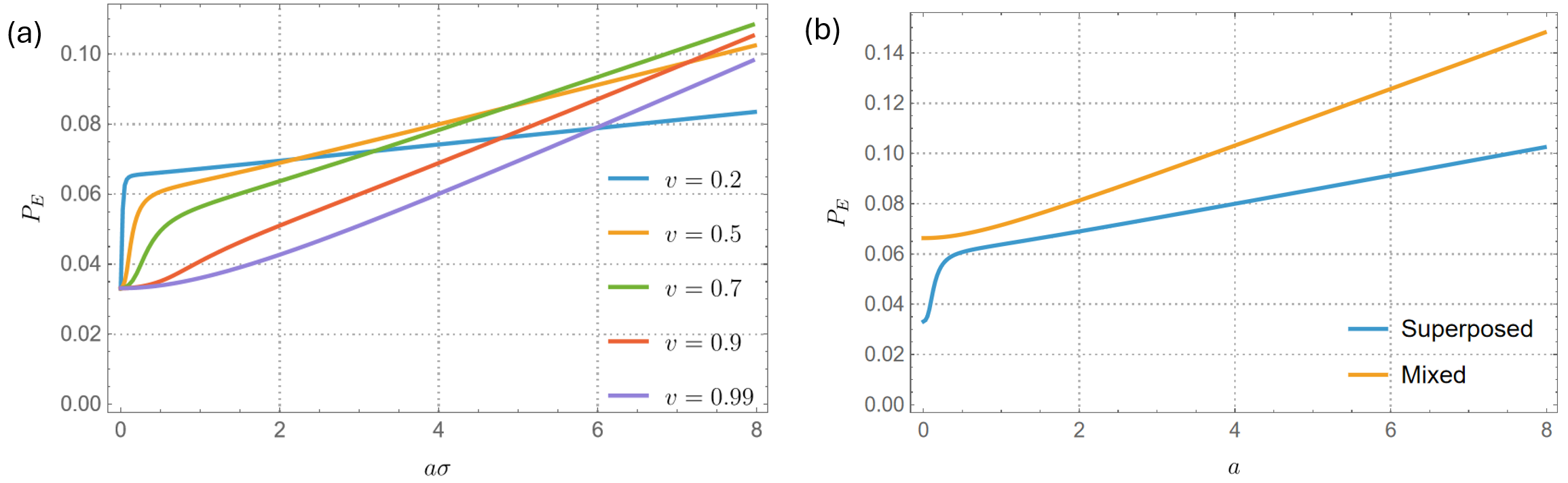}
\caption{
(a) (Blue) The transition probability of a detector in a superposition of a static point and an orbiting circle versus the centripetal acceleration of the orbit. (Orange) The transition probability for an equivalent mixed state system of identical trajectories. Here $\Omega\sigma=0.1$ and $v=0.5$. (b) The transition probability of a detector in a  superposition of a static point and an orbit versus the acceleration of the orbit. Here $\Omega\sigma=0.1$.
}
\label{fig:Static_P_vs_a_combined}
\end{figure*}

If instead the tangential velocity of the trajectory is kept constant and the acceleration is allowed to vary, we obtain the results shown in figure \ref{fig:Static_P_vs_a_combined}. Unlike a classical circular trajectory, the response rate of a superposed detector  rapidly increases for low accelerations.

In the ultrarelativistic limit ($\gamma\rightarrow \infty$), for any fixed $\tau$, the $\gamma\tau'$ dominates the off-diagonal Wightman function  
\begin{align}\label{ptultra}
    W(\tau',\tau)\approx -\frac{1}{4\pi^2\gamma^2}\frac{1}{(\tau'-i\tilde{\varepsilon})^2-v^2/a}\, ,
\end{align}
where $\tilde{\varepsilon}=\varepsilon/\gamma$. Notice that the real part of the poles coincides with the acceleration timescale $v^2/a$. Note also that the limit $v\rightarrow 1$ is taken at fixed $a$. For a Gaussian switching with (fixed) $\sigma$ much larger than $v^2/a$, the interference contribution to the transition probability as $\gamma\rightarrow\infty$ is dominated by the response function
\begin{align}
    \mathcal{F}(\Omega)\approx-\frac{\Omega}{2\pi\gamma^2}\text{sinc}\left(\frac{\Omega v^2}{a}\right)\Theta(-\Omega)\, ,
\end{align}
which is strongly suppressed due to the factor of $\gamma^2$ in the denominator. However, for an interaction of infinite duration with constant switching function, the approximation \eqref{ptultra} breaks down, as the $\tau$ integration eventually extends to the point where $\gamma \tau'-\tau\approx \gamma \tau'$ is violated, for any fixed $\gamma$.

In the nonrelativistic limit ($\gamma\approx 1$), the off-diagonal Wightman function becomes approximately stationary:
\begin{align}
    W(\tau,\tau')\approx -\frac{1}{4\pi^2}\frac{1}{(\tau-\tau'-i\varepsilon)^2-R^2}\, .
\end{align}
Keeping the same (constant) switching, the corresponding response function is
\begin{align}\label{central_nonrel}
    \mathcal{F}(\Omega)\approx-\frac{\Omega}{2\pi}\text{sinc}(\Omega R)\Theta(-\Omega)\, .
\end{align}
In this limit, the off-diagonal response \eqref{central_nonrel} also coincides with that of a static detector in a superposition of two locations separated by a distance $R$, as one might expect. 
\begin{figure}[ht]
\centering
\includegraphics[width=\columnwidth]{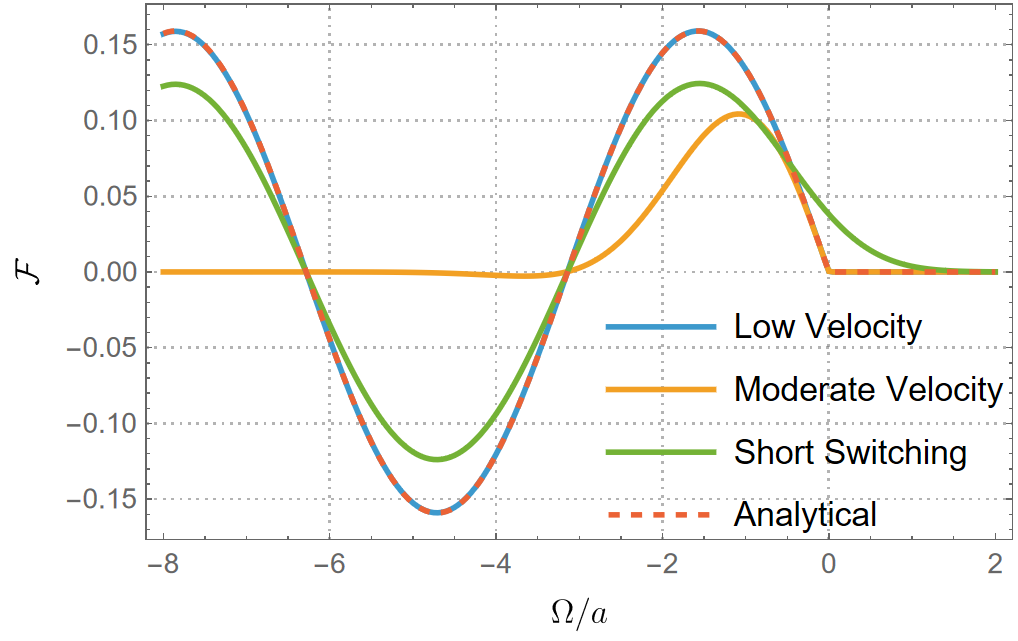}
\caption{
    The off-diagonal response function of a detector in a superposition of a static point and an orbiting circle versus the energy gap of the detector, normalized by the centripetal acceleration. We set $\gamma=1.0001$ to be in the low velocity regime (blue curve) and $\sigma a=100$ is taken to be large to remove as much of the effects of the switching function as possible. The moderate velocity curve (orange) is plotted for $\gamma = 1.01$, and the short switching curve (green) is plotted for $\sigma=1$. We see close agreement between the low velocity numerical result and the analytical approximation \eqref{stackedultra0}
    (red dashed curve).
}
\label{fig:Static_LowVel_analyticalcomparison}
\end{figure}
\begin{figure}[t]
\centering
\includegraphics[width=\columnwidth]{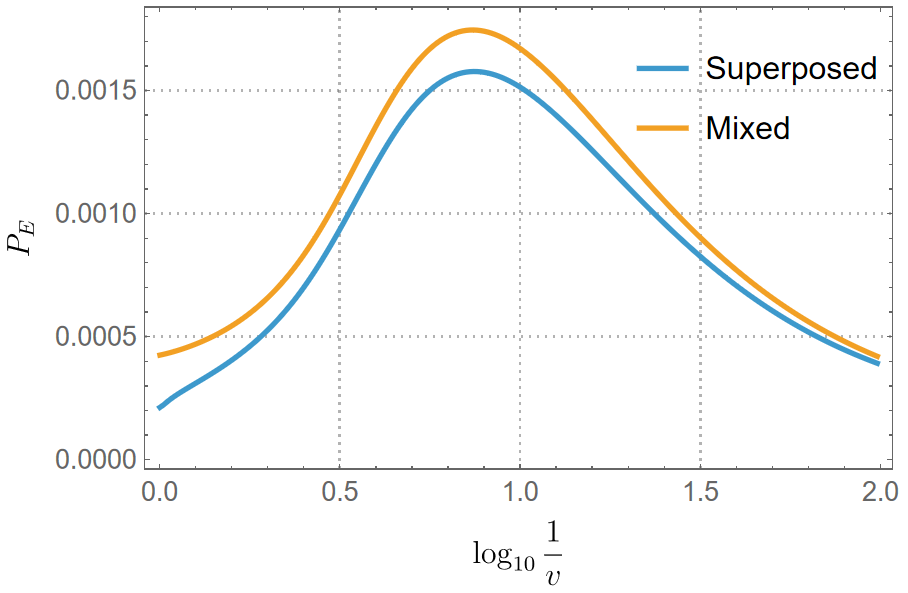}
\caption{
    (Blue) The transition probability of a detector in a superposition of concentric circles for $\Omega\sigma = 2$. (Orange) The transition probability for an equivalent mixed state system of identical trajectories. The acceleration of the outer trajectory is held fixed at $a\sigma=1$ and the inner trajectory orbits at the same rotational rate, but with half the radius of the outer trajectory. The response of an equivalent mixed state system is shown for comparison.
}
\label{fig:Conc_mixedcomparison}
\end{figure}
\begin{figure}
\centering
\includegraphics[width=\columnwidth]{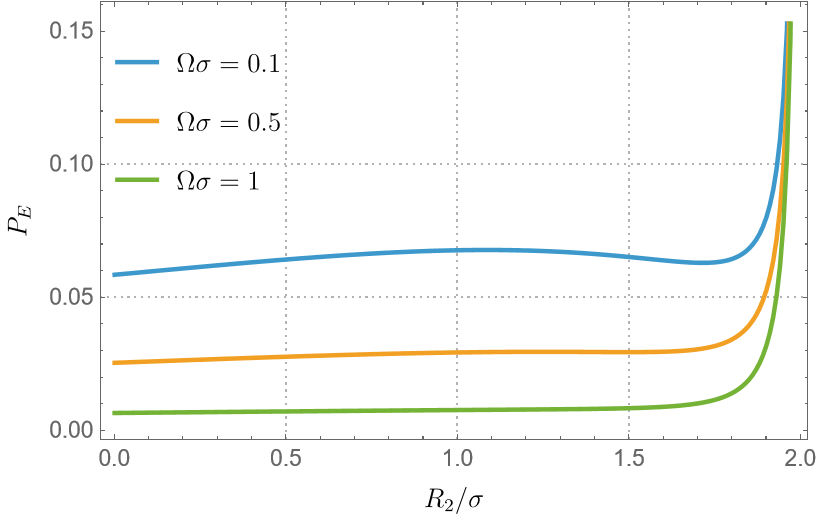}
\caption{
    The transition probability of a detector in a  superposition of concentric circles versus the radius of one of the circular trajectories. Here $\omega\sigma=1/2$ and the radius of one circle is held constant at $R_1/\sigma=1$.
}
\label{fig:Conc_P_vs_L}
\end{figure}

\subsection{Planar concentric circles}

Similarly, we can define a detector as being in a superposition of two concentric circular trajectories,
\begin{align}
    \sx_1(\tau)
    &=
        \enclose{
            \gamma_1 \tau, 
            R_1\cos (\omega_1 \gamma_1 \tau),
            R_1\sin (\omega_1 \gamma_1 \tau),
            0
        }\,\\
    \sx_2(\tau)
    &=
        \enclose{
            \gamma_2 \tau, 
            R_2\cos (\omega_2 \gamma_2 \tau),
            R_2\sin (\omega_2 \gamma_2 \tau),
            0
        }\,,
\end{align}
which corresponds to (d) in Figure \ref{fig:trajectories}. The Wightman function then becomes
\begin{widetext}
\begin{multline}
    W\ts{c}(s,p) = -\frac{1}{\pi^2}\bigg[\enclose{p(\gamma_1 - \gamma_2)+s(\gamma_1 + \gamma_2)}^2 - 4R_{1}^2 - 4R_{2}^2\\
    + 8R_{1}R_{2}\cos\benclose{\frac{1}{2}\enclose{s(\gamma_1\omega_1 + \gamma_2\omega_2) + p(\gamma_1\omega_1 - \gamma_2\omega_2)}}
    \bigg]^{-1}.
    \label{eq:concentric_wightman}
\end{multline}
Note that this expression reduces to equation \eqref{eq:diagonal_wightman} in the case when $R_2 = R_1$, and reduces to equation \eqref{eq:static_wightman} when $R_2$ goes to zero. 
\end{widetext}

The two trajectories possess different radii and can in general possess different angular velocities and centripetal accelerations. For expediency  we select trajectories such that their angular velocities are identical, meaning $\omega_1 = \omega_2$. This reduces the size of the parameter   space whilst accommodating expected physical limitations of the system at play; we imagine an experimental setup where the same mechanism drives the rotation for both possible trajectories of the detector. This also implies that the detector experiences different accelerations depending on the trajectory.

Figure \ref{fig:Conc_mixedcomparison} shows the effect of the superposition for a concentric set of trajectories. Much like the static/orbiting case, the superposition has a damping effect on the detector's transition probability. 
Dependence of the response over
the entire spectrum of  radii ratios 
  is shown in figure \ref{fig:Conc_P_vs_L}. At $R_2/\sigma=0$ we recover  the initial response for a static detector superposition. As the inner circle expands the response grows until it attains a local maximum when the radii are equal. The response then declines slighlty, but as  the velocity of the outer trajectory  approaches the speed of light 
 ($R_2/\sigma \to 2$) 
the response increases drastically.

\begin{figure}
\centering
\includegraphics[width=\columnwidth]{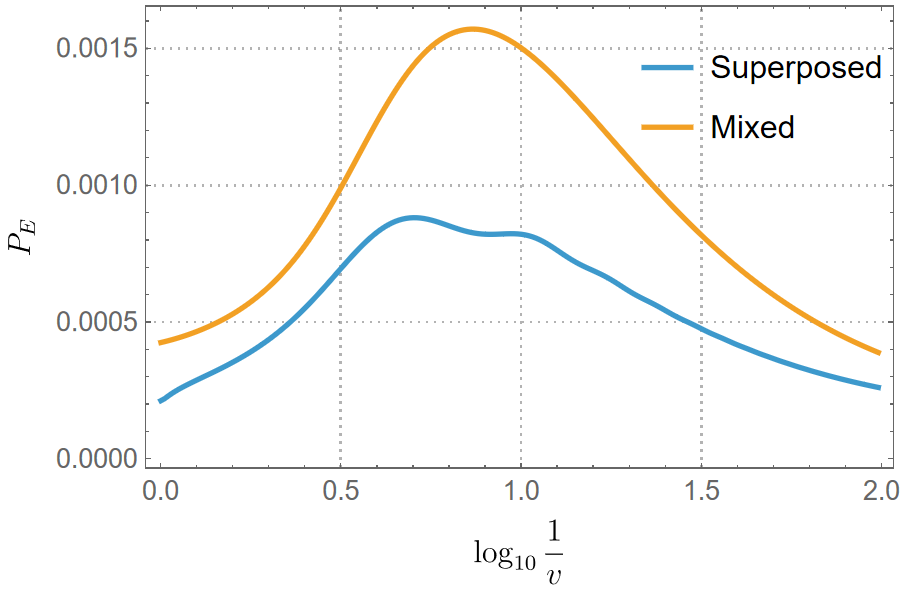}
\caption{
    (Blue) The transition probability of a detector in a superposition of vertically displaced concentric circular trajectories versus the tangential velocity of the outer circle. (Orange) The transition probability for an equivalent mixed state system of identical trajectories. Here $\Omega\sigma=2$, $H/\sigma=1$, and the centripetal acceleration of the circles is $a\sigma=1$. The inner circle has half the radius of the outer circle.
}
\label{fig:CZgap_mixedcomparison}
\end{figure}

\begin{figure}
\centering
\includegraphics[width=\columnwidth]{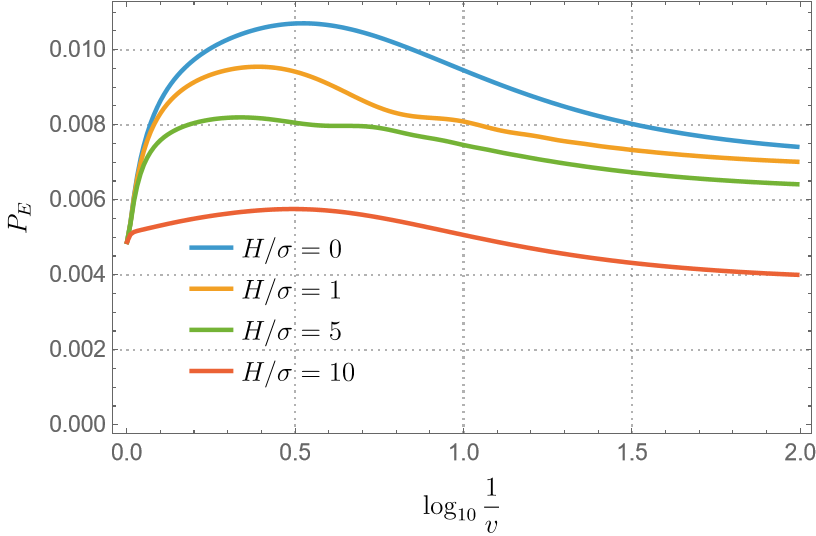}
\caption{
    The transition probability of a detector in a  superposition of vertically displaced concentric circles versus tangential velocity, $v$. Here $a\sigma=1$, $\Omega\sigma=0.1$ and the inner circle has $2/3$ the radius of the larger circle.
}
\label{fig:CZgap_P_vs_logb}
\end{figure}

The concentric configuration for the trajectories can also be vertically offset. The denominator of equation \eqref{eq:concentric_wightman} is modified by subtracting $4H^2$ from the other terms in the square brackets. For a general look at the Wightman function for every configuration, see appendix \ref{sec:general_wightman}. 
We display   the results  in figure \ref{fig:CZgap_mixedcomparison}. The vertical displacement of the circles reintroduces the oscillation of the response over a range of velocities. Figure \ref{fig:CZgap_P_vs_logb} presents the extent of the effect of displacement, showing how the appearance of oscillations is highly dependent on the magnitude of displacement in relation to the time scale of the interaction.

\section{Effective Temperature}
\label{sec:Teff}

Another metric by which we can measure the response of the system is through its temperature. In quantum field theory we typically describe the temperature of a system based on the Kubo-Martin-Schwinger (KMS) condition \cite{Fewster.Waiting.Unruh}. In practice, this entails that the detector response function satisfies the detailed balance relation,
\begin{align}\label{Tkms}
    e^{\Omega/T_{KMS}}=\mathcal{F}(-\Omega)/\mathcal{F}(\Omega)\, ,
\end{align}
which is valid for stationary trajectories and sufficiently large interaction times $\sigma$. In these cases the response function is directly proportional to the transition rate. It is only in situations where the KMS temperature $T_{KMS}$ is independent of the energy gap $\Omega$ that the detector response can be considered exactly thermal. 

In more general contexts, particularly situations with sufficiently narrow switching functions that prevent asymptotic thermalization, one can still define an effective temperature using the ratio of transition probabilities, rather than transition rates:
\begin{align}\label{Teff}
    e^{\Omega/T_{eff}}=P_E(-\Omega)/P_E(\Omega).
\end{align}
This measure of an effective temperature provides an intuitive estimator for the field's thermal behavior when we are looking at finite interactions like those we have modeled; in these cases, the effective temperature $T_{eff}$ can depend on the energy gap $\Omega$. For stationary scenarios in the limit of constant-switching interactions of infinite duration, the effective temperature \eqref{Teff} approaches the KMS temperature \eqref{Tkms}, and varies only weakly with $\Omega$. Here we examine the effective temperature for all of the above configurations of trajectories.

For the case of two trajectories in coherent superposition, one finds
\begin{align}\label{eq:Teff_superposed}    e^{\Omega/T_{eff}}=\frac{\left(P_{11}+P_{12}+P_{21}+P_{22}\right)|_{-\Omega}}{\left(P_{11}+P_{12}+P_{21}+P_{22}\right)|_{\Omega}}\, ,
\end{align}
whereas for an incoherent mixture we have
\begin{align}    e^{\Omega/T_{eff}}=\frac{\left(P_{11}+P_{22}\right)|_{-\Omega}}{\left(P_{11}+P_{22}\right)|_{\Omega}}\, .
\end{align}
This corresponds to the effective temperature of an ordinary detector in circular motion when the trajectories are identical and merely displaced. In those cases, the effective temperature corresponds to the actual KMS temperature of the field.
\begin{figure*}[ht]
\centering
\includegraphics[width=\textwidth]{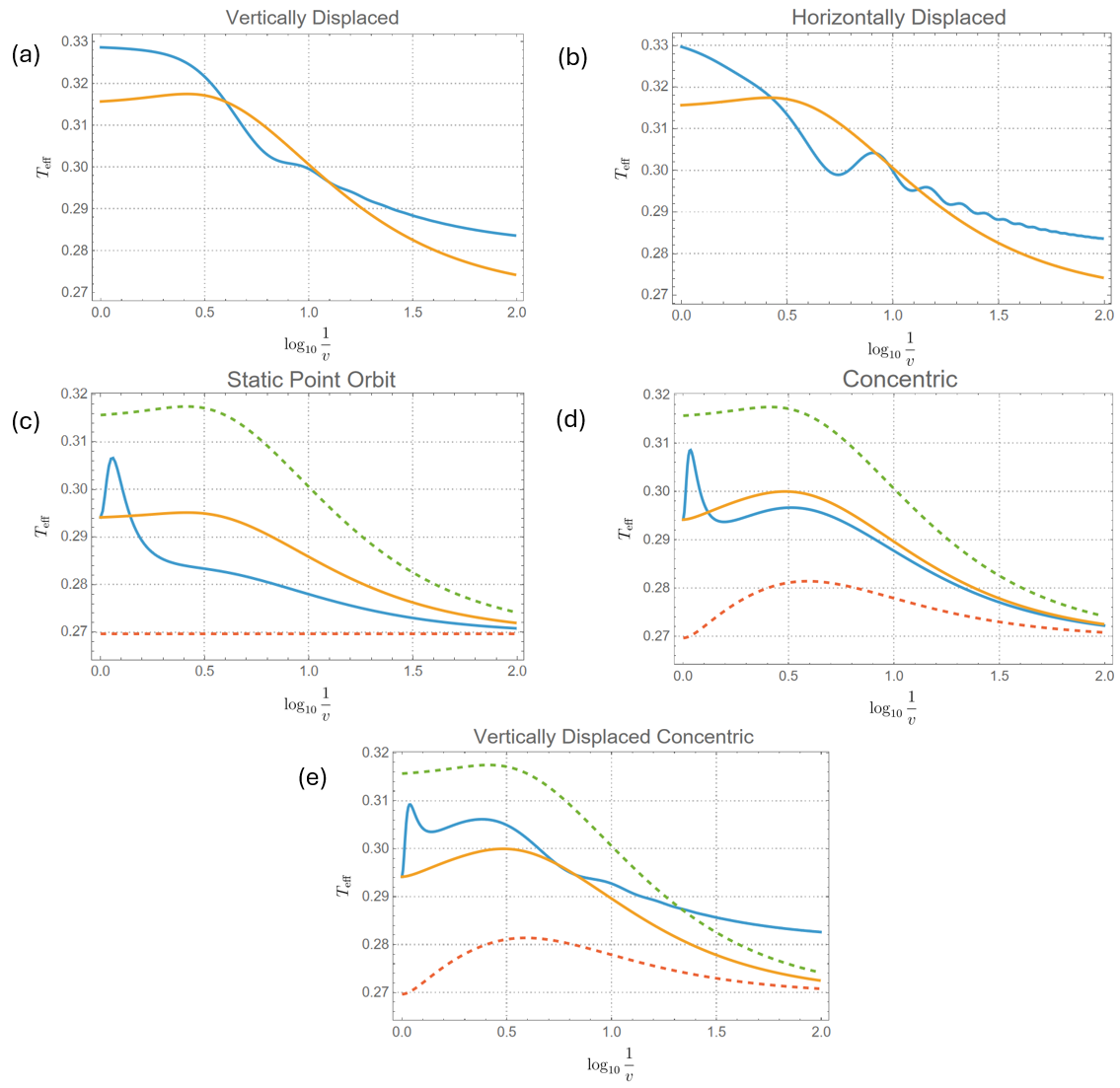}
\caption{
The effective temperature of a detector in  superposition (blue) and in a corresponding mixed state (orange)
for various detector configurations. 
The temperatures of the larger (green) and smaller (Red) of the two circles are given by dashed lines where applicable.    $\Omega\sigma=1$ for all   scenarios. \\ (a) Vertically displaced circles with $H/\sigma=1$ and $a\sigma=1$. (b) Horizontally displaced with $L/\sigma=1$ and $a\sigma=1$. (c) A static detector and an orbiting circle with $a\sigma=1$. (d) Concentric circles with $\omega_2=\omega_1$ and $R_2 = R_1/2$.  In the cases where the parameters of trajectories are different, the tangential velocity is taken to be that of the larger of the two circles, with $a\sigma=1$. (e) Vertically displaced concentric circles with $\omega_2=\omega_1$, $R_2 = R_1/2$; the tangential velocity is that of the larger of the two circles and $a\sigma=1$ for the larger circle as well.
}
\label{fig:Teff_comparison}
\end{figure*}
Figure \ref{fig:Teff_comparison} shows a comparison between coherent superpositions and incoherent mixtures for all of the configurations of trajectories that we have looked at. The two cases for displaced identical circles, (a) and (b) in the figure, show similar features. Both systems exhibit oscillations in temperature over a range of velocities, much like the transition probability, with horizontal displacement generating much stronger interference effects due to the superposition. They also possess higher temperatures than the non-superposed case as the detector approaches the speed of light, which differs from all other examined cases where the temperatures converged for extremely high velocities. In each case where the two superposed trajectories had different physical parameters, the temperature rises sharply for high velocities before rapidly converging back down to the same temperature as the incoherent mixture.

Similarly, the temperature for the superposed systems converges to that of the equivalent mixtures at low velocities only for configurations in which the trajectories are centered on the same point, as seen in figure \ref{fig:Teff_comparison} (c) and (d) for the static point and concentric circles respectively. Low linear velocities correspond to rapidly revolving small circles when we hold the acceleration constant, meaning that as the velocity becomes very small the trajectories of the detector become more and more spatially isolated from one another. At sufficiently low velocities, the range of motion from one of the superposed trajectories is not large enough to have a significant correlation with the other; instead the separation of the circles now dominates over the radius of circular motion.   Figure \ref{fig:Teff_comparison} (a), (b), and (e) show how displacement between the center of the circles causes the effective temperature of the detector to be significantly larger at lower velocities, and a direct comparison of the numerical results for both vertical and horizontal displacement show that they converge to the same value indicating that the orientation of the circles becomes irrelevant at that point.

\begin{figure*}[t]
\centering
\includegraphics[width=\textwidth]{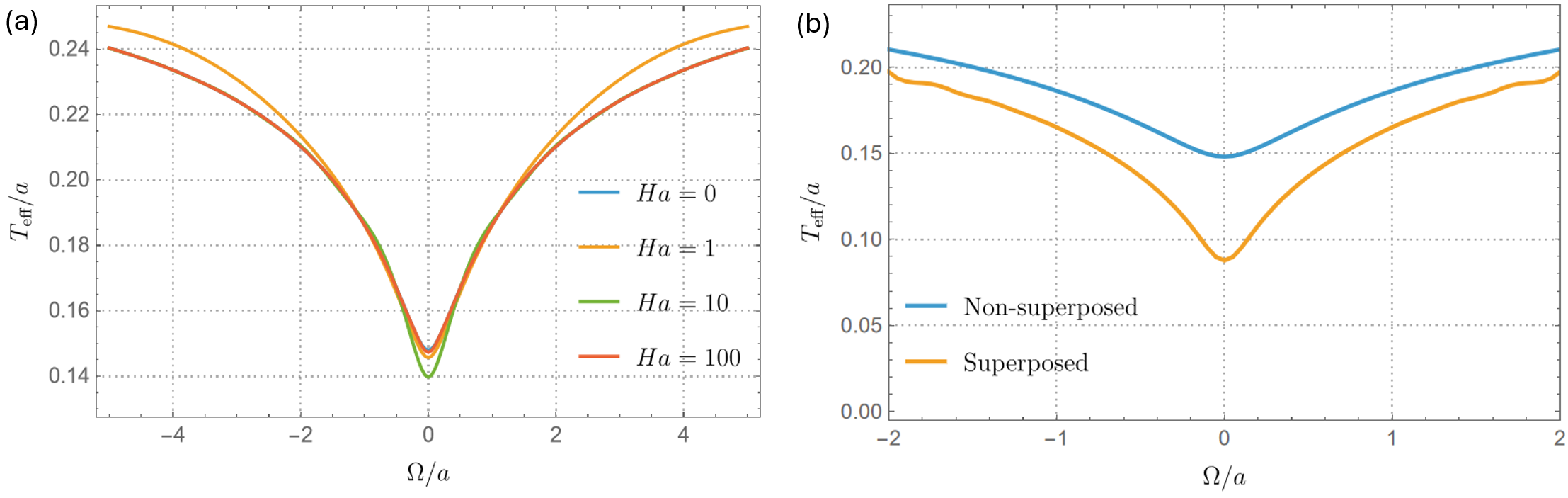}
\caption{
    (a) The effective temperature of a detector in a superposition of vertically stacked circular trajectories versus the energy gap of the detector, normalized by the centripetal acceleration. The results are shown for a range of vertical displacements, with $Ha=0$ corresponding to the result for a detector not in superposition. (b) The effective temperature of a detector in a superposition of horizontally displaced circular trajectories versus the energy gap of the detector, normalized by the centripetal acceleration. Here $Ha=1$. In both cases, $v=\sqrt{0.99}$ and $\sigma a=10$ is taken to be large in order to facilitate thermalization.
}
\label{fig:Combined_Teff_vs_Omega}
\end{figure*}

\subsection{Effective Thermalization}\label{sec:effective_thermal}

The results presented thus far have dealt entirely with systems where the interaction time is relatively short and finite. We are also interested in examining the thermal properties of the superposed detector for long interaction times where the detector has the opportunity to properly come to equilibrium. One particularly important result for circular trajectories is given by Unruh \cite{Unruh1998}, where the temperature of the detector is shown to be linear in the acceleration and varies slightly based on the energy gap. A natural line of inquiry is then to ask how such a relation changes for a superposed system.

Let's restrict our attention to a stationary case - say, vertically stacked circles of equal radius as described in section \ref{sec:vert_circ}. Using equation \eqref{eq:Teff_superposed},
we seek to plot $T_{eff}/a$ as a function of $\Omega/a$. The various terms involved are given by equation \eqref{eq:TransitionP_ij}.
Commensurate with \cite{Unruh1998}, we can set $\gamma=10$, which corresponds to $v=\sqrt{0.99}$. The results are shown in figure \ref{fig:Combined_Teff_vs_Omega} (a) for several different displacements. 

We first observe that the ordinary temperature relation is given by the $Ha=0$ line and is almost entirely obscured by the $Ha=100$ red line. For a large enough separation between the circles, the temperature relation converges exactly back into the 
non-superposed
scenario   because the correlations are no longer able to have any meaningful effect over such distances. However the dependence of the temperature on $H$ is not simple, as shown by the yellow and green lines. The detector can either become hotter for larger energy gaps, or colder for lower energy gaps depending on the magnitude of the distance between the trajectories in superposition.



If we instead focus on an alternate scenario where the detector is in a superposition of a static location and orbit (i.e. scenario (c) in Figure \ref{fig:trajectories}), we can perform the exact same analysis. In figure \ref{fig:Combined_Teff_vs_Omega} (b), we observe that such a superposition reduces the temperature of the detector (especially for detectors with small energy gaps), and also increases the variation of the temperature with energy gap. Unfortunately, practical limitations in the numerical methods prevented us from calculating the temperature for even larger values of $\Omega/a$ and different configurations as the computation quickly becomes unstable.


\section{Discussion}
\label{sec:Disc}


We have examined the response of an Unruh-DeWitt detector placed in a superposition of two separate circular trajectories in (3+1) dimensional Minkowski spacetime. In section \ref{sec:circ}, the possible configurations for these trajectories were broken down into four general cases: (a) vertically displaced, (b) horizontally displaced, (c) a static central point with surrounding circular trajectory, and (d) concentric circles. The results for a combination of the first and final cases were also examined. Both cases (a) and (b) involving displaced circular trajectories displayed oscillations in the transition probability for the detector over a range of linear velocities at constant acceleration, with both the strength and nature of the interference effects being highly sensitive to the magnitude of the displacement in relation to the switching time. In contrast, cases (c) and (d) did not display such oscillatory effects due to interference; instead the transition probability was heavily dampened based on the difference in parameters between the trajectories.

The effective temperature of each case was also examined in section \ref{sec:Teff}. In the absence of a definable KMS temperature for our system, an effective temperature measure was taken to be an estimator of the general thermal properties of the detector. The effective temperature of the detector in each case displayed much of the same interference effects as before, such as oscillations in temperature for displaced trajectories. In all cases, a superposed detector possessed a higher temperature at extremely high velocities. However, for all cases where the trajectories were synchronized in their rotations between trajectories with different physical parameters, the temperature converged with the mixed state case when approaching the speed of light. At the low velocity limit, only trajectories with non-displaced centers converged to the same temperature as the mixture.

In section \ref{sec:effective_thermal}, we calculated the effective temperature as a function of the energy gap for scenarios (a) and (c) (i.e. concentric, vertically-stacked trajectories and a static central point with surrounding circular trajectory, respectively), in the regime of wide switching functions compared to the acceleration timescale. We found that interference between trajectories had only a minor effect on the effective temperature in case (a), while in case (c) there was a much more pronounced effect: the temperature was significantly lower than the non-superposed case and had a greater variation with energy gap. 

All of the results shown here have assumed that the control state for the superposition of the detector was measured in the same state as it was initialized in. In general, the final state of the control can be conditional on being measured in any particular state
\begin{equation}
    \ket{\alpha} = \frac{1}{\sqrt{N}}\sum_{i}^2e^{-i\varphi_i}\ket{i},
\end{equation}
with the caveat that the interference terms of the transition probability possess a coefficient that suppresses the interference terms \cite{Foo:2020xqn}. Equation \ref{eq:Psum} becomes
\begin{equation}
    P_E = \frac{\lambda^2}{4}\enclose{\sum_{i}^2 P_{ii} + \cos\enclose{\delta\varphi}\sum_{i\neq j}P_{ij}},
    \label{eq:Psum_coherence}
\end{equation}
where $\delta\varphi = \varphi_1 - \varphi_2$. In the interest of maximizing the effects of the superposition on the detector's response, the outcome was presumed to be conditional on a final state measurement that was identical to the initial one such that $\delta\varphi=0$. Any particular determination of the final measurement of the control represents an outcome within a continuous range of behaviour between the superposed and mixed state cases shown throughout our results.

The results presented in this work have potential to provide experimental guidance for an analogue realization of superpositions of circular trajectories, building on the recent proposals to implement the circular motion Unruh effect in ultracold atom systems \cite{Analog1,Analog2,Analog4} and to test superpositions of detector trajectories \cite{twoplusone_preprint}. These experimental proposals exploit a mathematical analogy between a relativistic Klein-Gordon field and either density fluctuations in a pancake-like Bose-Einstein condensate \cite{Analog2,Analog1,twoplusone_preprint} or height fluctuations in a thin-film of superfluid helium \cite{Analog4}, with a continuum of Unruh-DeWitt detectors simulated by a highly-focused laser probe. In these experimental proposals, however, the effective Klein-Gordon field is $(2+1)$-dimensional, so one would expect slight differences in the predicted behaviour \cite{UnruhTemps}. Still, our results suggest in principle that interference effects for superposed circular trajectories should be observable in an analogue experiment. We found that such interference effects are generally prominent in the regime of narrow to moderate switching functions, realizable experimentally with pulsed laser probes \cite{Analog3}. Of particular interest is the regime of effectively constant switching functions, realizable with continuous-wave lasers, in scenario (c): a superposition of a static central point with a surrounding circular trajectory. This experimental configuration would only be a modest extension of the current proposals for analogue implementations of the circular motion Unruh effect.


\section{Acknowledgements} CG acknowledges support provided by the European Research Council under the Consolidator Grant COQCOoN (Grant No. 820079). This work was supported in part by the Natural Sciences and Engineering Research Council of Canada. 

For the purpose of open access, the authors have applied a CC BY public copyright licence to any Author Accepted Manuscript version arising.  

\bibliography{refs.bib}

\begin{appendix}

\section{General Wightman Function}
\label{sec:general_wightman}

The off diagonal Wightman functions given in the main body of the text, such as equations \eqref{eq:WightmanVertical} or \eqref{eq:static_wightman} for example, are specific to the particular configurations of superposed trajectories in their respective sections. It is of course possible to derive a generalized expression for the Wightman function of superposed circles in (3+1) dimensions, which we give here. For this scenario we allow the circles to have completely independent sets of parameters $\omega$ and $R$, and any form of displacement given by a difference in $x$ and $y$ coordinates by $L$ and $H$ respectively to coincide with the previous examples. The only stipulation we place is that both circles lie flat in the $x$-$y$ plane such that the detector never moves through the $z$ dimension in any given trajectory. The Wightman function for the interference terms is then given by
\begin{widetext}
\begin{equation}
    W(\tau,\tau') = -\frac{1}{4\pi^2}\frac{1}{(\gamma_1\tau - \gamma_2\tau')^2 + A_1 - A_2 + 2R_1R_2\cos\enclose{\gamma_1\omega_1\tau - \gamma_2\omega_2\tau'} - H^2 - L^2 - R_1^2 - R_2^2}.
\end{equation}
\end{widetext}
Here $A_1 = 2LR_1\cos\enclose{\gamma_1\omega_1\tau}$ and $A_2 = 2LR_2\cos\enclose{\gamma_2\omega_2\tau'}$. For the purposes of our numerical predictions, this expression proves to be especially cumbersome to use. It proves far more fruitful to simplify the expression through careful selection of parameters before performing any calculations, but it is given here for completeness.

\section{Numerical Details}
This section is dedicated to computing the transition probability and effective temperature for a UDW detector in a superposition of two trajectories. This is accomplished through numerical integration of \eqref{eq:TransitionP_ij} by inserting the appropriate Wightman function for the given configuration. Where possible, we take advantage of the stationarity of the trajectories to use \eqref{eq:StationaryP_ij} instead for the calculation.

For the purposes of computation, we calculate $P_E$ in units of $\lambda^2$ and $\sigma$ is taken to be of unit value. The response function of the detector is a function of $\Omega$, as well as $\omega$ and $R$ for both trajectories. The parameters for each circular trajectory are taken to be identical for the sake of simplicity unless otherwise stated. The rotational variables $\omega$ and $R$ can be replaced with the associated linear variables $a$ and $v$ where convenient, or in particular we can note that for circular trajectories specifically $1/v$ is equivalent to the torsion $b$ as well.

We wish to observe the effect of the superposition of trajectories as it compares to the response of an ordinary detector-field interaction. It is helpful then, to consider the non-superposed mixed state of a detector with a probability of being found in one trajectory or another,
\begin{equation}
    \rho = \frac{1}{N}\sum \ket{j}\bra{j}.
    \label{eq:Mixedstate}
\end{equation}
Following a very similar derivation of the excitation probability as before, we can find that only the $P_{jj}$ terms in \eqref{eq:Psum} are left. Note that this means for any configuration that has two circles with equivalent parameters (i.e. the displaced identical circles), the mixed state result corresponds exactly with the ordinary response of a single detector traveling along a circular trajectory, as one would expect since the relative positions of the circles has no effect in this case.

In order to numerically integrate the transition probability for a given system, steps must be taken to avoid running into divergences present on the real line due to poles present in the Wightman function. To that end, we calculate such quantities using contour integration where each pole is avoided using infinitesimally small semicircular offsets into the complex plane. Since our switching functions go to zero very quickly, we can approximate the full integral by integrating a specific range determined by the characteristic time of the gaussian. We can avoid tediously tracking down poles on the real line and setting up complicated contours of integration by instead making use of residue theorem to integrate along a closed contour including the real line for which we know no poles exist, as long as the enclosed space also contains no poles. For convenience we can shift our integral a small amount $\epsilon$ into the complex plane and integrate the contour that way without poles, including the small vertical connections to the real line at the endpoints to form a closed box contour.

For numerical integrals involving multiple variables, such as the nonstationary configurations, we need only shift from real integration to a complex contour for one of the two variables. With a suitable choice for $\epsilon$, all poles on the real axes can be avoided without resorting to complex integration in both variables. When calculating the integrals, we make the change of variables to $s=\tau_1-\tau_2$ and $p=\tau_1+\tau_2$ and allow $s$ to be the variable over which we integrate via a contour. 

There is also a choice as to whether $\epsilon$ can be taken as positive or negative, shifting the integral into the positive or negative imaginary half of the complex plane. This choice is dictated by Jordan's lemma, which prescribes the direction taken in order to ensure convergence between the numerical approximation and the true infinite integral along the real line. It is worth noting that since the imaginary exponential responsible for dictating this contains only $\Omega$ and the difference between the two $\tau$, the offset between $P_E(\Omega)$ and $P_E(-\Omega)$ is determined exactly by pole contributions. 

\end{appendix}

\end{document}